%Paper: gr-qc/9303027
%From: Alan Rendall <RENDALL@sbitp.itp.ucsb.edu>
%Date: Wed, 24 Mar 1993 18:27 PST

\magnification=1200
\hfuzz=2 pt
\def\d{\partial}
\def\R{{\bf R}}
\def\tr{{\rm tr}}
\def\ref#1{\lbrack #1\rbrack}
\def\next{\hfil\break\noindent}

\def\Quadrat#1#2{{\vcenter{\hrule height #2
  \hbox{\vrule width #2 height #1 \kern#1
    \vrule width #2}
  \hrule height #2}}}
\def\halmos{\ \ $\Quadrat{4pt}{0.4pt}$}
\font\title=cmbx12
{\title
\centerline{The Newtonian limit for asymptotically flat solutions}
\centerline{of the Vlasov-Einstein system\footnote*{{\rm Part of this
work was done during a visit to the Department of Physics, Syracuse
University, Syracuse NY 13244-1130, USA}}
}}
\vskip 2cm\noindent
Alan D. Rendall

\vskip 10pt\next
Max-Planck-Institut f\"ur Astrophysik
\next
Karl-Schwarzschild-Str. 1
\next
8046 Garching bei M\"unchen
\next
Germany
%\next
%FAX: 89 32993235
\next
E-mail: anr@cvxastro.mpa-garching.mpg.de

\vskip 2cm
%{\baselineskip 20pt
\noindent
{\bf Abstract}

It is shown that there exist families of asymptotically flat solutions
of the Einstein equations coupled to the Vlasov equation describing a
collisionless gas which have a Newtonian limit. These are sufficiently
general to confirm that for this matter model as many families of this
type exist as would be expected on the basis of physical intuition. A
central role in the proof is played by energy estimates in unweighted
Sobolev spaces for a wave equation satisfied by the second fundamental
form of a maximal foliation.

%\vskip .5cm
%\noindent
%Running title: Newtonian limit of the Vlasov-Einstein system

%\noindent
%Mathematics Subject Classification: 83C05

\vfil\eject

\noindent
{\bf 1. Introduction}

It is a well known empirical fact that in many situations a general
relativistic description of the motion of self-gravitating matter
can be replaced to a good approximation by a non-relativistic,
Newtonian one. In the usual formulation of Newtonian gravity the
interaction is described by a single scalar function, the Newtonian
potential. The relation of this to general relativity, where
the fundamental object is a Lorentz metric, is obscure. The basic
idea required to understand this relation mathematically was provided
by Cartan\ref4. He showed that Newtonian theory can be formulated in
such a way that the basic object is an affine connection whose
non-zero components are components of the gradient of the Newtonian
potential. The role of the potential itself is then merely that of
providing a convenient representation of this connection in certain
coordinate systems. It was realised by Friedrichs\ref{12}\ that the
natural way to connect the two theories is to require that the
Levi-Civita connection of the spacetime metric go over in the limit
as the speed of light $c$ goes to infinity into the connection
defined by Cartan. Since then many authors have extended this work
on the relations between the equations of the two theories and
the physical interpretations of their solutions. This knowledge has
been systematised in the frame theory of Ehlers (see \ref{11}\ and
references therein). What has been achieved is to set up a precise
definition of the Newtonian limit of general relativity which encodes
that which is desirable on physical grounds. The major open question
is whether this definition is compatible with the Einstein equations
in the sense that there exists a sufficiently large class of solutions
which satisfy all the axioms. The purpose of this paper is to answer
this question in the affirmative.

The case of the Newtonian limit which is of most physical interest is
that of an isolated system. This is expressed mathematically by
restricting attention to asymptotically flat solutions of the
Einstein equations. (The case of cosmological solutions, which is
also of considerable interest, will not be treated here.) It is of
prime importance to have results which do not only handle the vacuum
Einstein equations since in that case only the trivial Newtonian
solution (i.e. empty space) could be expected to arise as a limit
of singularity free asymptotically flat spacetimes. It is at this
point that the first serious difficulty is encountered. The most
obvious type of matter to take would be one or more bodies made of
fluid or an elastic solid. However only very limited results exist
on the initial value problem for self-gravitating bodies of this type
\ref{20}. It is for this reason that the matter model chosen here is a
collisionless gas described by the Vlasov equation. It is known that
the local in time initial value problem for the Vlasov-Einstein system
is well posed for a class of initial data which allows spatially
localised matter distributions\ref5. A large part of what follows does
not crucially depend on any property of the particular matter model
chosen beyond the fact that the local in time initial value problem
is well posed. There is, however, one step where a special property of
the Vlasov equation is used, namely in the last paragraph of section 3.
In order to generalise the results of this paper to a different matter
model it would be essential to find a replacement for the argument of
that paragraph or to modify the structure of the main proof
significantly to avoid the need for that argument.

The next difficulty which hampers the development of rigorous theorems
on the Newtonian limit is that this limit is singular in the sense that
the Einstein equations, which are essentially hyperbolic, go over into
the Poisson equation, which is elliptic. The hyperbolic nature of the
Einstein equations is related to the propagation of gravitational waves.
There is one special case, namely the case of spherical symmetry, where
gravitational radiation is absent. This leads to a simplification of
the problem and the Newtonian limit of spherically symmetric
asymptotically flat solutions of the Vlasov-Einstein system was handled
in \ref{19}.

When spherical symmetry is not assumed the problem of the singular
limit has to be faced and to see which way to go it is useful to
consider a simpler analogue of the Vlasov-Einstein system where a
limiting situation occurs which is rather similar. This is the
Vlasov-Maxwell system whose quasi-static limit has been considered in
\ref1,\ref8\ and \ref{22}. Of these papers the one which is of most
relevance here is that of Degond\ref8. He treats the limit using the
fact that in the energy estimates for the Maxwell equations the terms in
the equations which blow up as $c\to\infty$ make no contribution. The
solutions discussed belong to a Sobolev space on each slice of constant
time. For the Einstein equations this does not hold. An asymptotically
flat metric falls off only as $r^{-1}$ as $r\to\infty$ on a
spacelike slice and the positive mass theorem implies that any
attempt to impose faster fall-off excludes all but the trivial solution.
Thus the metric does not belong to a Sobolev space. The
usual way to get around this is to replace the ordinary Sobolev space
by a weighted one. Unfortunately it is easily seen that such a
replacement destroys the property used by Degond that singular terms
drop out of the energy estimates. In the following this difficulty is
circumvented by using a formulation of the Einstein equations due to
Christodoulou and Klainerman\ref6. There the only object which is
determined by solving a non-trivial hyperbolic equation is the second
fundamental form, which does lie in an ordinary Sobolev space. The
$r^{-1}$ part of the metric is generated by solving an elliptic
equation.

Now that the strategy has been outlined, the main theorem will be
stated. The notation is as follows: $g_{ab}$ is the induced metric
on the leaves of a maximal foliation of spacetime, $k_{ab}$ is the
second fundamental form of this foliation, $\phi$ is the lapse
function, $\Gamma^\alpha_{\beta\gamma}$ are the Christoffel symbols
of the spacetime metric and $f$ is the phase space density of particles.
The spacetime metric is of the form
$$ds^2=-\phi^2dt^2+g_{ab}dx^adx^b.\eqno(1.1)$$
The notions of regular initial data and regular solutions appearing in
the statement will be defined in Sect. 5. The exact interpretation of
the order symbols used will be given at the end of Sect. 6. The
parameter $\lambda$ corresponds physically to $c^{-2}$.
\vskip .25cm
\noindent
{\bf Theorem 1.1} {\it Let $(g_{ab}^0(\lambda), k_{ab}^0(\lambda),
f^0(\lambda))$ be a parameter-dependent initial data set for the
Vlasov-Einstein system which is regular of order $s$ for some $s\ge6$
and satisfies the constraints and the maximal slicing condition.
Suppose that as $\lambda\to 0$:

\noindent
(i) $g_{ab}^0(\lambda)=\lambda\delta_{ab}+O(\lambda^{3/2})$

\noindent
(ii) $k_{ab}^0(\lambda)=O(\lambda^{3/2})$

\noindent
(iii) $\d_t k^0_{ab}(\lambda)=O(\lambda^{3/2})$

\noindent
(iv) $f^0(\lambda)=f_N^0+o(1)$

\noindent
for some $f_N^0$.
Then a solution
$(g_{ab}(\lambda),k_{ab}(\lambda),\phi(\lambda),f(\lambda))$
of the Vlasov-Einstein system, which is regular of order $s$ and induces
the given initial data on the hypersurface $t=0$, exists on a
$\lambda$-independent time interval $\lbrack0,T)$ and has the
properties:

\noindent
(i) $g_{ab}(\lambda)=\lambda\delta_{ab}+o(\lambda)$

\noindent
(ii) $k_{ab}(\lambda)=o(\lambda)$

\noindent
(iii) $\phi(\lambda)=1-U\lambda+o(\lambda)$ for some $U$

\noindent
(iv) $\Gamma^a_{00}(\lambda)=-\delta^{ab}\nabla_bU+o(1)$

\noindent
(v) all other components $\Gamma^\alpha_{\beta\gamma}$ are
$o(1)$

\noindent
(vi)$f(\lambda)=f_N+o(1)$ for some $f_N$

\noindent
Moreover $f_N$ and $U$ solve the Vlasov-Poisson system with initial
datum $f_N^0$.}

\vskip .25cm\noindent
In assumption (iii) in the hypotheses of this theorem the time
derivative is to be calculated using the Einstein evolution equation
(2.4) and the function $\phi$ in that equation is to be got by solving
the lapse equation (2.12). This assumption may seem unnatural but it is
essential. Its significance will be discussed further in Sect. 7.

It is appropriate at this point to mention some recent work related to
the present paper. Fritelli and Reula\ref{13}\ have suggested an
interesting approach to proving convergence of solutions of the Einstein
equations in the Newtonian limit on a spatially bounded region.
Lottermoser\ref{15}\ has proved the existence of rather general families
of solutions of the Einstein constraint equations having a Newtonian
limit. It was necessary to prove some new results on existence of
families of solutions of the constraints in the present paper since
Lottermoser's method is not suitable for producing solutions which
satisfy a prescribed gauge condition (e.g. the maximal slicing condition
used in the following). It is also of interest that the solutions whose
existence is demonstrated in the present paper include ones which do not
belong to the class produced in \ref{15}. This is because the basic
object $Z_{ab}$ used there and supposed to behave regularly as
$\lambda\to 0$ diverges in general in the Newtonian limit for the data
constructed here.

The paper is organised as follows. In Sect. 2 the form of the Einstein
equations used in \ref6\ is discussed and it is shown how the
parameter $\lambda$ can conveniently be introduced into it. In the third
and fourth sections estimates are derived for the Vlasov and Einstein
equations respectively. In Sect. 5 these are used to prove the local
existence of a solution on a $\lambda$-independent interval and the
remainder of Theorem 1.1 is proved in Sect. 6. The existence of a
large class of regular initial data is demonstrated in Sect. 7. Two
appendices are concerned with some elliptic theory and estimates for
modified Sobolev spaces which are needed in the body of the paper.

\vskip .5cm
\noindent
{\bf 2. Derivation of the reduced equations}

Consider first the 3+1 form of the Einstein equations with zero shift.
The constraints are
$$\eqalignno{
R-|k|^2+(\tr k)^2&=16\pi\phi^{-2}T_{00},&(2.1)        \cr
\nabla^ak_{ab}-\nabla_b\tr k&=-8\pi\phi^{-1}T_{0b},&(2.2)}$$
and the evolution equations are
$$\eqalignno{&\d_t g_{ab}=-2\phi k_{ab},&(2.3)        \cr
&\d_t k_{ab}=-\nabla_a\nabla_b\phi+\phi(R_{ab}+\tr kk_{ab}-2k_{ac}k^c_b
\cr
&-8\pi T_{ab}-4\pi\phi^{-2}T_{00}g_{ab}+4\pi Tg_{ab}).&(2.4)}$$
Here $R_{ab}$ is the Ricci tensor of $g_{ab}$. The objects $T_{00}$,
$T_{0a}$ and $T_{ab}$ are components of the matter tensor and $T$ is
obtained by taking the trace of $T_{ab}$ with the metric $g_{ab}$.
Define
$$\eqalignno{
A&=\tr k,&(2.5)                        \cr
B&=R-|k|^2,&(2.6)                      \cr
C_a&=\nabla^b k_{ab}-{1\over2}\nabla_a(\tr k),&(2.7)        \cr
D_{ab}&=\phi^{-1}\d_t k_{ab}+\phi^{-1}\nabla_a\nabla_b\phi-R_{ab}
+2k_{ac}k^c_b.&(2.8)}$$
It will also be useful to have the modified quantities
$$\eqalignno{
\tilde B&=B-16\pi\phi^{-2}T_{00},&(2.9)                \cr
\tilde C_a&=C_a+8\pi\phi^{-1}T_{0a},&(2.10)            \cr
\tilde D_{ab}&=D_{ab}+8\pi T_{ab}+4\pi (\phi^{-2}T_{00}-T)g_{ab}.&(2.11)
}$$
In the following the maximal slicing condition $A=0$ will be used.
Under that assumption equations (2.1), (2.2) and (2.4) are equivalent to
$\tilde B=0$, $\tilde C_a=0$ and $\tilde D_{ab}=0$ respectively. If
$A=0$ then the lapse equation
$$\Delta\phi=(|k|^2+4\pi\phi^{-2}T_{00}+4\pi T)\phi\eqno(2.12)$$
is satisfied. Following \ref6\ it can be shown that (2.1)-(2.4)
together with $A=0$ imply a wave equation for $k_{ab}$.
$$-(\phi^{-1}\d_t)^2 k_{ab}+\Delta k_{ab}=N_{ab}+\tau_{ab},\eqno(2.13)$$
where
$$\eqalignno{
&\tau_{ab}=8\pi\left\lbrack {1\over2}\phi^{-3}\dot T_{00}g_{ab}
-\phi^{-1}(\nabla_a T_{0b}+\nabla_b T_{0a})+\phi^{-1}\dot T_{ab}
\right.                \cr
&\left.-{1\over2}\phi^{-1}\dot Tg_{ab}+\phi^{-2}(T_{0a}\nabla_b\phi+
T_{0b}\nabla_a\phi)-(\phi^{-2}T_{00}-T)k_{ab}-\phi^{-4}\dot\phi T_{00}
g_{ab}\right\rbrack,&(2.14)                                \cr
&N_{ab}=L_{ab}-H_{ab},&(2.15)                              \cr
&\phi^2L_{ab}=\nabla_a\nabla_b\dot\phi-\phi^{-1}\dot\phi
\nabla_a\nabla_b\phi-\dot\Gamma^c_{ab}\nabla_c\phi
+2\phi(k_a^c\d_tk_{bc}+k_b^c\d_tk_{ac})                    \cr
&+4\phi^2k_{ac}k^{cd}k_{bd},&(2.16)                        \cr
&\phi H_{ab}=\phi I_{ab}+\nabla^c\phi(2\nabla_ck_{ab}-\nabla_ak_{bc}
-\nabla_bk_{ac})-\nabla_a\phi\nabla^ck_{bc}
-\nabla_b\phi\nabla^ck_{ac}                                \cr
&-\nabla_c\nabla_b\phi k^c_a-\nabla_c\nabla_a\phi k^c_b
+\Delta\phi k_{ab}
&(2.17)                                                    \cr
&\phi I_{ab}=-3(R_{ac}k^c_b+R_{bc}k^c_a)+2g_{ab}(k^{cd}R_{cd})+k_{ab}R.
&(2.18)}$$
The reduced system of Einstein equations which will be used consists
essentially of (2.3), (2.12) and (2.13). Unfortunately the occurrence
of the Ricci tensor in $I_{ab}$ causes trouble with the existence
theory and so it will be replaced by using the relation
$$R_{ab}=\phi^{-1}\d_t k_{ab}+\phi^{-1}\nabla_a\nabla_b\phi+2k_{ac}
k^c_b+8\pi T_{ab}+4\pi(\phi^{-2}T_{00}-T)g_{ab}\eqno(2.19)$$
which is equivalent to $\tilde D_{ab}=0$. The following lemma was proved
in the vacuum case in \ref6.

\noindent
{\bf Lemma 2.1.} {\it Let $(g_{ab},k_{ab},\phi)$ be a solution of
(2.3), (2.12) and (2.13) (with (2.19) having been substituted into
(2.13) to eliminate $R_{ab}$). Then if the data $(g_{ab}(0),
k_{ab}(0),\d_t k_{ab}(0))$ are such that $A$, $\tilde B$, $\tilde C_a$
and $\tilde D_{ab}$ vanish for $t=0$ and if $\nabla_\alpha T^{\alpha
\beta}=0$ then $A$, $\tilde B$, $\tilde C_a$ and $\tilde D_{ab}$
vanish everywhere so that $(g_{ab},k_{ab},\phi)$ defines a solution
of the Einstein equations for which the hypersurfaces $t$=const. are
maximal.}

\noindent
{\it Proof} This will only be sketched since it is very similar to the
vacuum case. Equations (2.3), (2.12) and (2.13) imply (with
$F=B+\tr D$):
$$\eqalignno{
&\phi^{-1}\d_t A=\tilde F:=F-4\pi\phi^{-2}T_{00}-4\pi T,&(2.20)      \cr
&\phi^{-1}\d_t F=\Delta A-4\phi^{-1}\nabla^a\phi C_a
+\phi^{-1}(R-\Delta\phi)A                              \cr
&\qquad -8\pi\lbrack -2\phi^{-1}\nabla^a T_{0a}+2\phi^{-2}\nabla^a\phi
T_{0a}-1/2\phi^{-1}\dot T-2k^{ab}T_{ab}        \cr
&\qquad -3\phi^{-4}\dot\phi T_{00}+(-\phi^{-2} T_{00}
+T)A+3/2\phi^{-3}\dot T_{00}\rbrack&(2.21)  \cr
&\phi^{-1}\d_t\tilde C_a=\nabla^b\tilde D_{ab}-1/2\nabla_a(\tr \tilde D)
+\phi^{-1}\nabla^b\phi\tilde D_{ab}-1/2\phi^{-1}\nabla_a\phi\tilde F
\cr &\qquad
-\phi^{-1} A\nabla^b\phi k_{ab}-\nabla^b Ak_{ab}+8\pi\phi^{-1}T_{0a}A
&(2.22)                                     \cr
&\phi^{-1}\d_t\tilde D_{ab}=\nabla_a\tilde C_b+\nabla_b\tilde C_a
&(2.23)}$$
In the derivation of (2.22) the vanishing of the 4-dimensional
divergence of $T^{\alpha\beta}$ has been used. The latter can also be
used to simplify the equation obtained from (2.21) by substituting
for $F$ and $C_a$ in terms of $\tilde F$ and $\tilde C_a$. Equations
(2.20)-(2.23) imply wave equations of the form
$$\eqalignno{
&-(\phi^{-1}\d_t)^2 A+\Delta A=M,&(2.24)                           \cr
&-(\phi^{-1}\d_t)^2 \tilde F+\Delta \tilde F=M^\prime,&(2.25)      \cr
&-(\phi^{-1}\d_t)^2 \tilde C_a+\Delta \tilde C_a=M_a^{\prime\prime},
&(2.26)}$$
where $M$, $M^\prime$ and $M^{\prime\prime}$ are linear expressions in
the quantities $A$, $\nabla A$, $\nabla^2 A$, $\tilde F$,
$\nabla\tilde F$, $\phi^{-1}\d_t\tilde F$, $\tilde C$, $\nabla\tilde C$,
$\phi^{-1}\d_t\tilde C$, $\tilde D$, $\nabla\tilde D$. The proof of the
lemma can now be completed by deriving an inequality of the form
$${\cal E}(t)\le K\int_0^t{\cal E}(s)ds\eqno(2.27)$$
where
$$\eqalign{
{\cal E}&=\int_{\R^3}(|A|^2+|\nabla A|^2+|\tilde F|^2+|\nabla\tilde F|^2
+|\phi^{-1}\d_t\tilde F|^2                       \cr
&\qquad +|\tilde C|^2
+|\nabla\tilde C|^2+|\phi^{-1}\d_t\tilde C|^2+|\tilde D|^2) dV_g,}
\eqno(2.28)$$
and $dV_g$ is the volume element on $\R^3$ associated with the metric
$g_{ab}$. This would be a straightforward consequence of the usual
energy inequalities for wave equations if it were not for the occurrence
of the quantities $\nabla\tilde D$ and $\nabla^2 A$ on the right hand
side of the equations. Note that these do not appear in the definition
of $\cal E$. In fact $\nabla\tilde D$ only occurs in $M^{\prime\prime}$
and $\nabla^2 A$ only in $M^\prime$ and the problem can be overcome as
follows. One of the terms which needs to be estimated is schematically
of the form $\int_0^t\lbrack
\int_{\R^3}(\d_t\tilde C\nabla\tilde D)(s)\rbrack
ds$. Integrating by parts in time converts this into the sum of a
spacetime integral and a boundary contribution on the hypersurface
labelled by $t$. The spacetime integral can now be handled by a partial
integration in space. A partial integration in space should also be
applied to the boundary term. It is then schematically of the form
$\int_{\R^3}(\nabla\tilde C\tilde D)$. This can be estimated by an
expression of the form
$$K\left(\eta\int_{\R^3}|\nabla\tilde C|^2
+\eta^{-1}\int_{\R^3}|\tilde D|^2\right)$$
where $K$ is a constant and $\eta$ may be chosen to be any positive
real number. Choosing it so that $K\eta\le1/2$ we can absorb the first
term into $\cal E$. To handle the second term express it as the integral
from $0$ to $t$ of its time derivative and use equation (2.23). The
term containing $\nabla^2A$ can be estimated in an analogous way.
Applying Gronwall's inequality to (2.27) now completes the proof.
\halmos

Lemma 2.1 shows that providing we are dealing with a matter model
which guarantees that $\nabla_{\alpha}T^{\alpha\beta}=0$ (and this is
in particular true of matter described by the Vlasov equation) then
solving the reduced system consisting of (2.3), (2.12) and (2.13)
suffices to solve the Einstein equations. In this paper the unknowns
$(g_{ab},k_{ab},\phi)$ are time-dependent objects on $\R^3$.
Normally, if asymptotically flat situations are to be studied,
the boundary conditions $g_{ab}\to\delta_{ab}$ and $\phi\to1$ as
$|x|\to\infty$ will be imposed. In order to study the Newtonian limit
the first of these will be replaced by $g_{ab}\to \lambda\delta_{ab}$
where the parameter $\lambda$ corresponds to $c^{-2}$. The Newtonian
limit then corresponds to the limit $\lambda\to0$ and if this is to
be regular it will be the case that $g_{ab}=O(\lambda)$ and
$k_{ab}=O(\lambda)$ as $\lambda\to0$. In the terminology of the
frame theory (see \ref{11}) $g_{ab}$ is part of the temporal metric.
In view of this dependence on $\lambda$ it is convenient to use the
variables $\gamma_{ab}=\lambda^{-1}g_{ab}$ and $\kappa_{ab}=\lambda^{-1}
k_{ab}$. Then $\gamma_{ab}$ satisfies the standard boundary condition
that $\gamma_{ab}\to\delta_{ab}$ as $|x|\to\infty$. The basic equations
are
$$\eqalignno{
&\d_t\gamma_{ab}=-2\phi\kappa_{ab},&(2.29)           \cr
&\Delta_\gamma\phi=\lambda\lbrack |\kappa|^2+4\pi\phi^{-2}T_{00}+4\pi T
\rbrack \phi&(2.30)                                  \cr
&-(\phi^{-1}\d_t)^2\kappa_{ab}+\lambda^{-1}\Delta_\gamma\kappa_{ab}
=\lambda^{-1}(N_{ab}+\tau_{ab}).&(2.31)}$$
In equation (2.30) the norm of $\kappa_{ab}$ is defined by $\gamma_{ab}$
and not by $g_{ab}$.

\vskip .5cm
\noindent
{\bf 3. Estimates for the Vlasov equation}

For a discussion of the definition of the Vlasov equation in general
relativity see \ref{18}. Recall that the phase space density, which is
the unknown in this equation, is a real-valued function on the mass
shell $P$ i.e. the submanifold of the tangent bundle of spacetime
defined by the conditions $g_{\alpha\beta}p^\alpha p^\beta=-1$ and
$p^0>0$. The manifold $P$ can be coordinatized by the spacetime
coordinates $x^\alpha$ together with the spatial components $p^a$ of
the momentum. In the situation considered in this paper it is identified
in this way with $\R^6\times\lbrack 0,T)$. Using the 3+1 formalism
introduced in the previous section the explicit form of
the Vlasov equation is
$$\eqalign{&{\d f\over \d t}+\phi(1+\lambda|p|^2)^{-1/2}p^a{\d f\over
\d x^a}         \cr
&\qquad -\left\lbrack(1+\lambda|p|^2)^{1/2}\gamma^{ac}(\lambda^{-1}
\nabla_c\phi)
-2\phi\kappa^a_c p^c+\phi(1+\lambda|p|^2)^{-1/2}\Gamma^a_{bc}p^bp^c
\right\rbrack {\d f\over\d p^a}=0,}\eqno(3.1)$$
where $|p|^2=\gamma_{ab}p^ap^b$. Attention will be confined to initial
data for $f$ which are compactly supported in $\R^6$. Also $\lambda$
will be restricted to belong to the interval $(0,\lambda_0\rbrack$ for
some $\lambda_0>0$. No loss of generality results since it is only the
limiting behaviour as $\lambda\to 0$ which is of interest here.

In this section estimates for the solution of (3.1) will be obtained
in a fixed background geometry. It will be supposed that the quantities
$\Gamma^a_{bc}$, $\kappa_{ab}$, $\phi$ and $\lambda^{-1}\nabla_a\phi$
are continuous and bounded together with their first derivatives
with respect to the spatial coordinates $x^a$. Let $C_1$ denote a
common bound for these. It will furthermore be assumed that there exists
a positive constant $A$ such that $\phi^{-1}<A$ and
$A^{-1}\delta_{ab}\le\gamma_{ab}\le
A\delta_{ab}$. It follows that a similar estimate holds for
$\gamma^{ab}$. Equation (3.1) says that $f$ is constant along
characteristics (in the present context these are the lifts of geodesics
to the mass shell) and the assumptions on the geometry are enough to
guarantee the existence of these characteristics. Let $R(t)$ and $P(t)$
denote the maximum values of $|x|$ and $|p|$ respectively contained in
the support of $f$ at time $t$. Then (3.1) implies estimates of the
form $P(t)\le P(0)(1+Ct)e^{Ct}$ and $R(t)\le R(0)+\int_0^tP(s) ds$ where
the constant $C$ depends only on $C_1$ and $A$. Next the Sobolev norms
of $f$ will be estimated under the additional assumptions that
$\gamma_{ab}$ belongs to $L^\infty(\lbrack 0,T),K^s(\R^3))$ and that
$\kappa_{ab}$ and $\lambda^{-1}\nabla_a\phi$
belong to $L^\infty (\lbrack 0,T),H^s(\R^3))$. Here and in the following
$H^s(\R^n)$ denotes the standard Sobolev space of order $s$ and
$\|\ \|_{H^s}$ is the norm on that space. The $L^p$ norm is denoted
by $\|\ \|_p$. The space $K^s(\R^3)$ is defined to consist of those
functions $f$ in $L^\infty(\R^3)$ with $\nabla f\in H^{s-1}(R^3)$
with norm $\|f\|_{K^s}=\|f\|_\infty+\|\nabla f\|_{H^{s-1}}$.
Consider now the norm of $f(t)$ in $L^2(\R^6)$. Liouville's theorem
implies that the $L^2$
norm of $f(t)$ with respect to the geometrically natural volume form
is constant. On the other hand, under the assumptions already made
on the geometry, this volume form defines an equivalent $L^2$ norm
to that of the standard volume form on $\R^6$. Hence $\|f(t)\|_2\le C$
for a constant $C$ only depending on $A$ and $C_1$. To estimate
$\|f(t)\|_{H^s}$ for $s>0$ a method used in \ref8\ will be adopted. If
the Vlasov equation (3.1) is written schematically as $Xf=0$ then the
derivative $D^sf$ of order $s$ satisfies an equation of the form
$X(D^sf)=Q_s$ for a certain source term $Q_s$. At this stage it is
necessary to use the assumptions that $\gamma_{ab}$ belongs to
$L^\infty(\lbrack 0,T),K^s(\R^3))$ and that $\kappa_{ab}$ and
$\lambda^{-1}\nabla_a\phi$ belong to
$L^\infty(\lbrack 0,T),H^s(\R^3))$.
Let $C_2$ be a bound for their norms in these spaces. Note that if
$s\ge2$ the assumptions made up to now imply that
$\gamma^{ab}\in
L^\infty(\lbrack 0,T), K^s(\R^3))$ since $K^s(\R^3)$ is then a Banach
algebra. To estimate $Q_s$ the facts will be used that (see \ref{14},
\ref{16}\ or \ref{23}):
$$\eqalignno{
\|D^k(gh)\|_2&\le C(\|g\|_\infty\|h\|_{H^s}+\|g\|_{H^s}
\|h\|_\infty)&(3.2)                      \cr
\|D^k(gh)-gD^kh\|_2&\le C(\|Dg\|_\infty\|h\|_{H^{s-1}}+
\|g\|_{H^s}\|h\|_\infty)&(3.3)}$$
for $k\le s$.
These hold provided the norms on the right hand side of the inequalities
exist. They are valid for functions defined on the whole
of $\R^n$ or on a bounded domain. In this section the latter case is
the one which is relevant, the domain in question being an open subset
of $\R^6$ which contains the intersection of the support of $f$ with
each hypersurface of constant time. There results the estimate
$$\|Q_{s-1}\|_2
\le C\|f\|_{H^{s-1}}\ \ \ \ \ \ \ \ \ \ {\rm for}\  s\ge6.
\eqno(3.4)$$
Here $C$ depends only on $A$, $C_1$ and $C_2$ and the Sobolev
embedding theorem in $\R^6$ has been used. Combining (3.4),
the equation $X(D^sf)=Q_s$ and the inequality
$$\|f\|_{H^s}\le C(\|f\|_2+\|D^s f\|_2)\eqno(3.5)$$
shows that an integral inequality of the following form holds for
$s\ge6$.
$$\|f(t)\|_{H^{s-1}}^2\le\|f(0)\|_{H^{s-1}}^2
+C\int_0^t\|f(t^\prime)\|_{H^{s-1}}^2 dt^\prime,\eqno(3.6)$$
where $C$ only depends on $A$, $C_1$ and $C_2$. In fact $C_1$ can be
estimated in terms of $C_2$ and $\|\phi\|_\infty$ using the Sobolev
embedding theorem.

The matter tensor is defined by
$$T^{\alpha\beta}=-\int fp^\alpha p^\beta\phi|\gamma|^{1/2}
/p_0\ dp^1dp^2dp^3,\eqno(3.7)$$
where $\gamma$ denotes the determinant of $\gamma_{ab}$.
This is essentially the energy-momentum tensor. It is referred to
here as the matter tensor since it has been normalised so that in
the Newtonian limit $T^{00}$ becomes the mass density rather than
the energy density.
The $H^{s-1}$ norm of the integrand in (3.7) can be estimated by a
constant (depending only on $A$, $C_1$ and $C_2$) times the $H^{s-1}$
norm of $f$. Also for all functions $F(x,p)$ whose supports are
contained in a given compact set an inequality of the following form
holds.
$$\|\bar F\|_{H^{s-1}}\le C\| F\|_{H^{s-1}},\eqno(3.8)$$
where $\bar F(x)=\int F(x,p) dp$. Combining this information with (3.6)
gives
$$\|T^{\alpha\beta}(t)\|_{H^{s-1}}^2\le
C(\|f(0)\|_{H^{s-1}}^2+\int_0^t\|f(t^\prime)\|_{H^{s-1}}^2
dt^\prime).\eqno(3.9)$$
Suppose next that the time derivatives of $\gamma_{ab}$, $\kappa_{ab}$
and $\phi$ satisfy estimates of the same kind as already assumed for
the quantities themselves. Then an argument analogous to that above
leads to estimates of the form
$$\eqalignno{
&\|\d_t f(t)\|_{H^{s-2}}^2\le \|\d_t f(0)\|_{H^{s-2}}^2+C\int_0^t
\|f(t^\prime)\|_{H^{s-1}}^2+\|\d_t f(t^\prime)\|_{H^{s-2}}^2 dt^\prime,
&(3.10)                                                           \cr
&\|\dot T^{\alpha\beta}(t)\|_{H^{s-2}}^2\le C(\|\d_t f(0)\|_{H^{s-2}}^2
+\int_0^t \|f(t^\prime)\|_{H^{s-1}}^2+
\|\d_t f(t^\prime)\|_{H^{s-2}}^2 dt^\prime),&(3.11)}$$
where the constant $C$ now depends on $A$, $C_1$, $C_2$, $\|\dot\phi
\|_\infty$, the norm of the time derivative of $\gamma_{ab}$ in the
space $L^\infty(\lbrack 0,T),H^{s-1}(\R^3))$ and the norms of the time
derivatives of $\kappa_{ab}$ and $\lambda^{-1}\nabla\phi$ in
$L^\infty(\lbrack 0,T),H^s(\R^3))$.

The derivative $D^{s-1}(\d_t f)$ still needs to be estimated. To do
this, note that the Vlasov equation says that the solution is
constant along characteristics. Thus the value of $f$ at any point
is equal to the value of $f^0$ at the point where the characteristic
through that point intersects the initial hypersurface. Let
$(X(s,t,x,p), V(s,t,x,p))$ be the characteristic satisfying
$X(t,t,x,p)=(x,p)$ and $V(t,t,x,p)=(x,p)$. Then
$$f(t,x,p)=f^0(X(0,t,x,p),V(0,t,x,p)).\eqno(3.12)$$
Hence
$$\eqalign{
\d_t f(t,x,p)&=D_x f^0(X(0,t,x,p), V(0,t,x,p)) dX/ds(t,x,p)    \cr
&\qquad+D_p f^0(X(0,t,x,p), V(0,t,x,p)) dV/ds(t,x,p)}\eqno(3.13)$$
Now the quantities $dX/ds$ and $dV/ds$ are bounded in
$H^{s-1}(\R^6)$. Assuming that $f^0$ is in $H^s(\R^6)$ we see that
in order to estimate the $H^{s-1}$ norm of $\d_t f$ it suffices to
know the following two things. Firstly, the mapping taking $(t,x,p)$
to $X(0,t,x,p)$ (which is a $C^1$ diffeomorphism) is bounded in
$H^{s-1}$. Secondly, for $s$ sufficiently large (in the present case
$s>5$), the $H^{s-1}$ norm of the composition $g\circ h$ of a mapping
$g$ of class $H^{s-1}$ and a diffeomorphism $h$ of class $H^{s-1}$ can
be estimated in terms of the $H^{s-1}$ norms of $g$ and $h$. These
facts follow from the results of \ref{10}\ and \ref{17}. Thus the
$H^{s-1}$ norm of $\d_t f$ can be estimated by an expression of the
form $C\|f^0\|_{H^s}$ where $C$ depends only on the $H^{s-1}$ norm of
the coefficients in the Vlasov equation. An estimate for $\|\dot
T^{\alpha\beta}\|_{H^{s-1}}$ follows.

\vskip .5cm
\noindent
{\bf 4. Estimates for the Einstein equations}

First the form of $N_{ab}$ will be examined when it is written in
terms of $\gamma_{ab}$ and $\kappa_{ab}$. The indices of $\kappa_{ab}$
will be raised and lowered using $\gamma_{ab}$ and its inverse.
$$\eqalignno{
&\phi^2 L_{ab}=\nabla_a\nabla_b\dot\phi-\phi^{-1}\dot\phi
\nabla_a\nabla_b\phi
-\dot\Gamma^c_{ab}\nabla_c\phi                               \cr
&+2\lambda\phi
(\kappa_a^c\d_t\kappa_{bc}+\kappa_b^c\d_t\kappa_{ac})+4\phi^2
\lambda\kappa_{ac}\kappa^{cd}\kappa_{bd}&(4.1)               \cr
&\phi H_{ab}=\phi I_{ab}+\gamma^{cd}\nabla_c\phi(2\nabla_d\kappa_{ab}
-\nabla_a\kappa_{bd}-\nabla_b\kappa_{ad})                    \cr
&-\nabla_a\phi\gamma^{cd}\nabla_c\kappa_{bd}-\nabla_b\phi\gamma^{cd}
\nabla_c\kappa_{ad}-\nabla_c\nabla_a\phi\kappa^c_b-\nabla_c\nabla_b
\phi\kappa^c_a+\Delta_\gamma\phi\kappa_{ab}&(4.2)           \cr
&\phi I_{ab}=-3(R_{ac}\kappa^c_b+R_{bc}\kappa^c_a)+2\gamma_{ab}
(\kappa^{cd}R_{cd})+\kappa_{ab}\gamma^{cd}R_{cd}.&(4.3)}$$
In (4.3) it is still necessary to make the substitution
$$R_{ab}=\lambda\phi^{-1}\d_t\kappa_{ab}+\phi^{-1}\nabla_a\nabla_b\phi
+2\lambda\kappa_{ac}\kappa^c_b+8\pi T_{ab}+4\pi\lambda(\phi^{-2} T_{00}
-T)\gamma_{ab}.\eqno(4.4)$$
Suppose that a collection of quantities $(\gamma_{ab},\kappa_{ab},\phi,
T^{\alpha\beta})$ is given. These are not assumed to satisfy any
equations. If $\bar\gamma_{ab}$ is a solution of
$$\d_t\bar\gamma_{ab}=-2\phi\kappa_{ab},\eqno(4.5)$$
with initial datum $\gamma_{ab}(0)$ then an estimate of the form
$$\|\bar\gamma_{ab}(t)-\gamma_{ab}(0)\|_{H^s}^2
\le C\int_0^t \|\bar\gamma_{ab}(t^\prime)-\gamma_{ab}(0)\|_{H^s}^2
+\|\nabla\phi(t^\prime)\|_{K^{s-1}}^2
\|\kappa_{ab}(t^\prime)\|_{H^s}^2 dt^\prime\eqno(4.6)$$
holds for any $s\ge2$ provided the norms appearing all exist. This can
conveniently be proved using (A18). The time derivative of
$\bar\gamma_{ab}$ can be estimated similarly in terms of $\phi$,
$\kappa_{ab}$ and their time derivatives giving for any $s\ge3$ the
estimate
$$\eqalign{\|\d_t\bar\gamma_{ab}(t)\|_{H^{s-1}}^2 &\le
\int_0^t\|\d_t\gamma_{ab}(t^\prime)\|_{H^{s-1}}^2
+\|\nabla\phi(t^\prime)\|_{K^{s-1}}^2
\|\d_t\kappa_{ab}(t^\prime)\|_{H^{s-1}}^2 \cr
&\qquad+\|\nabla\dot\phi(t^\prime)
\|^2_{K^{s-1}}\|\kappa_{ab}(t^\prime)\|_{H^{s-1}} dt^\prime.}
\eqno(4.7)$$
Let $\bar\Gamma^a_{bc}$
denote the Christoffel symbols of $\bar\gamma_{ab}$. These satisfy
$$\d_t\bar\Gamma^a_{bc}=-\nabla_b(\phi k^a_c)-\nabla_c(\phi k^a_b)
+\nabla^a(\phi k_{bc}).\eqno(4.8)$$
In Sect. 5 this will be used to obtain a stronger estimate for
$\d_t\bar\Gamma^a_{bc}$ than could be derived from the estimate for
$\d_t\gamma_{ab}$. Next consider the following
equation which is closely related to (2.30).
$$\Delta_f\bar\phi=
-(\gamma^{ab}-\delta^{ab})\d_a\d_b\bar\phi-
\gamma^{ab}\Gamma^c_{ab}\nabla_c\bar\phi+\lambda\lbrack|\kappa|^2
+4\pi\phi^{-2}T_{00}+4\pi T\rbrack\bar\phi\eqno(4.9)$$
It is shown in the appendix that if $\gamma_{ab}$ is a metric such
that $\|\gamma_{ab}-\delta_{ab}\|_{K^{s-1}}
+\|\gamma_{ab}-\delta_{ab}\|_p$
is sufficiently small, if
the contents of the square brackets are in $L^1(\R^3)\cap H^{s-1}(\R^3)$
and if $\lambda$ is sufficiently small then (4.9) has a solution
$\bar\phi$ tending to 1 at infinity. Moreover if bounds for
$\|\gamma_{ab}-\delta_{ab}\|_{K^s}$, $\|\gamma_{ab}-\delta_{ab}\|_p$
and $\|\rho\|_1+\|\rho\|_{H^{s-1}}$
are given then a bound for
$\|\bar\phi\|_\infty+\lambda^{-1}\|\nabla\bar\phi\|_{H^s}$ is obtained.
Here $\rho$ denotes the expression in square brackets in (4.9).
Now the norms of $\rho$ appearing in this estimate
can be estimated in terms of the $H^s$ norm of $\kappa_{ab}$,
the $K^s$ norm of $\gamma_{ab}-\delta_{ab}$ and the $H^{s-1}$ norm of
$T^{\alpha\beta}$ together with a bound on the size of the support
of $T^{\alpha\beta}$. If equation (4.9) is differentiated
with respect to $t$ then the resulting equation can be used to obtain
bounds on $\d_t\bar\phi$ and its spatial derivatives in a manner similar
to the above.

The next equation for which estimates are needed is
$$-(\phi^{-1}\d_t)^2\bar\kappa_{ab}
+\lambda^{-1}\Delta_\gamma\bar\kappa_{ab}
=\lambda^{-1}(N_{ab}+\tau_{ab}).\eqno(4.10)$$
Suppose that a solution of (4.10) is given with initial data
$(\kappa_{ab}(0),\d_t\kappa_{ab}(0))$. The fundamental energy estimate
for the equation (4.10) is obtained by multiplying it by
$\gamma^{ac}\gamma^{bd}\d_t\bar\kappa_{cd}$ and integrating in space.
Define
$$E=\int_{\R^3}|\phi^{-1}\d_t\bar\kappa_{ab}|^2+\lambda^{-1}|\nabla
\bar\kappa_{ab}|^2+|\bar\kappa_{ab}|^2 dV_\gamma.\eqno(4.11)$$
Then the energy estimate takes the form
$$E(t)\le E(0)+C\left(\int_0^t E(t^\prime) dt^\prime+\int_0^t
\left|\int_{\R^3}
\lambda^{-1}\left\lbrack\gamma^{ac}\gamma^{bd}(N_{cd}+\tau_{cd})\d_t
\bar\kappa_{ab}\right\rbrack(t^\prime)dV_\gamma\right|
dt^\prime\right)\eqno(4.12)$$
where the constant $C$ depends on the norm in the space
$L^\infty(\lbrack 0,T), K^s(\R^3))$ of the quantity
$\gamma_{ab}-\delta_{ab}$ for $s\ge3$, the norm in the space
$L^\infty(\lbrack 0,T), H^s(\R^3))$ of the quantities
$\kappa_{ab}$ and $\lambda^{-1}\nabla_a\phi$, the $L^\infty$ norm of
$\d_t\gamma_{ab}$, the norm of $\lambda^{-1/2}\dot\Gamma^a_{bc}$ in the
space $L^\infty(\lbrack 0,T), H^{s-1}(\R^3))$, the $L^\infty$ norms of
$\phi$ and $\dot\phi$ and a constant $A$ such that $A^{-1}I\le
\gamma\le AI$ where $I$ is the identity matrix. This is proved in a
way which is standard for quasilinear hyperbolic equations (cf.
\ref{16}). The important point is that the only possible $\lambda$
dependence of the constant $C$ in (4.12) is through the quantities
$\lambda^{-1}\nabla\phi_n$ and $\lambda^{-1/2}\dot\Gamma^a_{bc}$
and it will be seen below that the latter quantities can be bounded in
terms of quantities which do not depend on $\lambda$.

To estimate the second term on the right hand side of (4.12) one could
try to use the estimate
$$\left|\int_{\R^3}S^{ab}\d_t\bar\kappa_{ab}\right|\le C\int_{\R^3}
|S|^2+|\d_t\kappa|^2\eqno(4.13)$$
for any tensor $S_{ab}$. This is sufficient for $N_{ab}$ and for most
of the terms in $\tau_{ab}$. However it is not sufficient to estimate
the terms involving spatial derivatives of $T^{\alpha\beta}$ without
resulting in a loss of differentiability in the iteration to be carried
out in Sect. 5. This difficulty can be overcome by the device of
doing a partial integration in time already used in the proof of
Lemma 2.1. The term to be estimated is schematically of the form
$\int_0^t\lbrack\int_{\R^3}(\d_t\kappa\nabla T)(t^\prime)\rbrack
dt^\prime$. Now integrate by parts in time. The resulting spacetime
integral can be handled by integration by parts in space. After partial
integration in space the boundary term is of the form $\int_{\R^3}
T\nabla\kappa$. Now estimate this by the sum of
a small constant times the $L^2$
norm of $\nabla\kappa$ and a large constant times the $L^2$ norm of
$T^{\alpha\beta}$. The former term can be absorbed into the energy
and the latter can be estimated by expressing it as the integral of
its time derivative from $0$ to $t$.
If it is known that $\bar\Gamma^a_{bc}$ is $O(\lambda^{1/2})$ then
the estimate for $\lambda^{-1}\int |\nabla_a\bar\kappa_{bc}|^2$
coming from (4.12) implies a similar one for
$\lambda^{-1}\int |\d_a\bar\kappa_{bc}|^2$.

Now consider the higher derivatives of
$\kappa_{ab}$. Differentiate (4.10) up to $s-1$ times (using partial,
not covariant, derivatives) and rearrange the result to give a
hyperbolic equation for $D^m\kappa$, $m\le s-1$. Let $E_m$ be defined
by replacing $\kappa$ in the definition of $E$ by $D^m\kappa$. An
inequality similar to (4.12) can then be obtained with $E$ replaced
by $E+\sum_{i=1}^{s-1} E_i$ and the term involving $N$ and $\tau$ being
replaced by terms involving derivatives of $N$ and $\tau$ up to order
$s-1$. Here a bound for the $H^{s-1}$ norm of
$\lambda^{-1/2}\bar\Gamma^a_{bc}$
goes into the constant appearing for the following reason. Many terms of
the
form $\lambda^{-1}D^{m_1}\gamma D^{m_2}\bar\kappa$ with $m_1, m_2\ge1$
occur
as source terms in the equation for $D^m\kappa$. Hence it is necessary
to have bounds for spatial derivatives of $\lambda^{-1/2}\gamma$. These
can be obtained from the equation
$$\d_a \gamma_{bc}=\gamma_{bd}\Gamma^d_{ac}
+\gamma_{cd}\Gamma^d_{ab}\eqno(4.14)$$
Note that the quantity $E+\sum_{i=1}^{s-1} E_i$ defines a
norm which is stronger than $\|\kappa_{ab}\|_{H^s}+\|\d_t\kappa_{ab}
\|_{H^{s-1}}$ because of the fact that it contains $\lambda^{-1}$.

\vskip .5cm
\noindent
{\bf 5. Existence on a uniform time interval}

In this section the existence statement of Theorem 1.1 will be proved.
First the definition of regular initial data must be given.

\noindent
{\bf Definition} An initial data set $(g^0_{ab}(\lambda),
k^0_{ab}(\lambda), f^0(\lambda))$ for the Vlasov-Einstein system
depending on a parameter $\lambda\in (0,\lambda_0\rbrack$ is called
{\it regular of order $s$} if for some $p<6$

\noindent
(i) $g^0_{ab}(\lambda)-\lambda\delta_{ab}$ belongs to the space
$L^p(\R^3)\cap K^{s+2}(\R^3)$ for each fixed $\lambda$

\noindent
(ii) $k^0_{ab}(\lambda)$ belongs to the space $H^s(\R^3)$ for each
$\lambda$

\noindent
(iii) $f^0(\lambda)$ belongs to $H^s(\R^6)$ and has compact
support for each $\lambda$

\noindent
It is of course assumed that $g^0_{ab}(\lambda)$ is a Riemannian metric
for
each $\lambda$.
\vskip .2cm

If a regular initial data set satisfies the conditions (i) and (ii)
occurring in the hypotheses of Theorem 1.1 then the quantities
$\gamma^0_{ab}=\lambda^{-1} g^0_{ab}$ and
$\kappa^0_{ab}=\lambda^{-1}k^0_{ab}$
satisfy the conditions
$$\eqalign{
&\gamma^0_{ab}=\delta_{ab}+O(\lambda^{1/2}),        \cr
&\kappa^0_{ab}=O(\lambda^{1/2}).}\eqno(5.1)$$
The $O$-symbol is to be understood in the sense of the function spaces
occurring in the above definition, both in (5.1) and in the hypotheses
of
Theorem 1.1. Consider now equation (2.30) on the initial hypersurface.
Using the definition (3.7) the expression in the square bracket in
(2.30)
can be brought into a form which does not contain $\phi$. The results of
the
appendix show that for $\lambda$ sufficiently small this equation has a
solution $\phi$ which tends to 1 at infinity and that $\nabla\phi$ is
$O(\lambda)$ as $\lambda\to 0$ in the space $H^s(\R^3)$.
It follows that
if $\d_t\kappa_{ab}$ is defined with the help of (2.4) then it belongs
to $H^s(\R^3)$ for each $\lambda$. Assumption (iii) of Theorem 1.1
implies that it is $O(\lambda^{1/2})$ in that space as $\lambda\to 0$.
In
this way data for the reduced system (2.29)-(2.31) can be constructed
from regular initial data for the Vlasov-Einstein system.

To prove the existence theorem an iteration will be set up. First
define $\gamma_0$, $\kappa_0$, $\phi_0$, $f_0$ by extending the
initial data $\gamma^0$, $f^0$ and the function $\phi^0$ in a
time independent manner and defining $\kappa_0(t)=\kappa^0+
(\d_t\kappa)^0t$. These functions do not satisfy equations
(2.29)-(2.31) but do satisfy the desired initial conditions. If now
$\gamma_n$, $\kappa_n$, $\phi_n$ and $f_n$ have been defined the next
iterate is obtained as follows. First solve (4.5) with $\phi$ and
$\kappa$ replaced by $\phi_n$ and $\kappa_n$ and $\gamma^0$ as initial
datum. Define $\gamma_{n+1}$ to be equal to the solution $\bar\gamma$.
This should only be done on an interval $\lbrack 0, T_n)$ short enough
so that $\gamma_{n+1}$ is a Riemannian metric. Next solve (4.9) with
$\gamma$, $\kappa$, $\phi$ and $f$ replaced by the corresponding
quantities with a subscript $n$ and define $\phi_{n+1}$ to be equal to
the solution $\bar\phi$. To ensure the existence of the solution it
may be necessary to reduce the size of $T_n$. The value of $\lambda_0$
must also be restricted in a way described below.
Now solve (4.10)
with $\gamma$ replaced by $\gamma_n$ and the quantities occurring in the
definitions of $N_{ab}$ and $\tau_{ab}$ replaced by the corresponding
quantities with subscript $n$. As initial data use $\kappa^0$ and
$(\d_t\kappa)^0$. Let $\kappa_{n+1}$ be equal to the solution
$\bar\kappa$ obtained. Finally, in order to obtain $f_{n+1}$ solve (3.1)
with $\gamma$, $\kappa$ and $\phi$ and $f$ replaced by $\gamma_n$,
$\kappa_n$, $\phi_n$ and $f_{n+1}$ respectively and initial datum $f^0$.

The next step is to show that this iteration is bounded in certain
function spaces and that the $T_n$ can be chosen so that $T_n\ge T$
for all $n$, where $T$ is a positive constant. Define
$$\eqalignno{
&a_n(t)=\max_{0\le m\le n}\sup_\lambda\left\{
\|\gamma_m(t)-\gamma_m(0)\|_{H^s}^2+\|\d_t\gamma_m(t)\|_{H^{s-1}}^2
+\|\kappa_m(t)\|_{H^s}^2\right.\cr
&\left.+\|\d_t\kappa_m(t)\|_{H^{s-1}}^2
+\|\lambda^{-1/2}\nabla\kappa_m(t)\|_{H^{s-1}}^2
+\|f_m(t)\|_{H^{s-1}}^2+\|\d_t f_m(t)\|_{H^{s-2}}^2\right\}&(5.2)
\cr
&b_n(t)=\max_{0\le m\le n}\sup_\lambda\left\{\|\phi_m(t)\|_\infty^2
+\|\lambda^{-1}\nabla\phi_m(t)\|_{H^s}^2+\|\dot\phi_m(t)\|_\infty^2
\right.  \cr
&\left.+\|\lambda^{-1}\nabla\dot\phi_m(t)\|_{H^s}^2
+\|\lambda^{-1/2}\dot\Gamma_m(t)\|_{H^{s-1}}^2+\|D^{s-1}(\d_t
f_m)(t)\|_2^2
\right\}&(5.3)}$$

\vskip 10pt\noindent
{\bf Lemma 5.1} {\it Let $a_n(t)$ and $b_n(t)$ be the functions on
$\lbrack 0,T_n)$ defined (5.2) and (5.3). Let $\lambda_0$ be a positive
constant. Define
$$\rho_n=|\kappa_n|^2+4\pi\phi_n^{-2}(T_{00})_n+4\pi T_n.\eqno(5.4)$$
Suppose that for some $p<6$ the inequalities
$\|\gamma_{ab}-\delta_{ab}\|_{K^{s-1}}+\|\gamma_{ab}-\delta_{ab}\|_p\le
C_1$
and $\lambda_0(\|\rho_n(t)\|_\infty+\|\rho_n(t)\|_{H^{s-1}})\le C_2$
hold on the whole interval $\lbrack 0,T_n)$, where the constants $C_1$
and $C_2$ are chosen so that Lemma A1 is applicable. Then there exists a
constant $C$, only depending on the initial data, and positive
real-valued functions $D$ and $D^\prime$ on $\R^2$ and $\R$ respectively
which are bounded on bounded subsets such that for all $n>1$:
$$\eqalignno{
&a_{n+1}(t)\le C+ D(\sup a_n, \sup b_n)
\int_0^t a_{n+1}(t^\prime)+b_n(t^\prime) dt^\prime
&(5.5)
\cr
&b_{n+1}(t)\le D^\prime(\sup a_n)&(5.6)}$$
}
{\it Proof} This is an application of the various estimates derived for
the Vlasov and Einstein equations in sections 3 and 4. Note first that
the required estimates for $\|\gamma_n(t)-\gamma_n(0)\|_{H^s}$ and
$\|\d_t\gamma_n(t)\|_{H^{s-1}}$ follow from (4.6) and (4.7)
respectively. The estimates for $\|\kappa_n\|_{H^s}$,
$\|\d_t\kappa_n\|_{H^{s-1}}$ and
$\|\lambda^{-1/2}\nabla\kappa_n\|_{H^{s-1}}$ are obtained from (4.12)
and the analogous estimate for $E+\Sigma_{i=1}^{s-1} E_i$ as described
towards the end of section 4. The inequalities (3.6) and (3.10) provide
the desired estimates for
$\|f_n\|_{H^{s-1}}$ and $\|\d_t f_n\|_{H^{s-2}}$. This completes the
proof of (5.5). The estimation of $\phi_{n+1}$ and its spatial
derivatives is accomplished by applying the information on the solution
of (4.9) given in section 4. The quantity
$\|\lambda^{-1/2}\dot\Gamma_n(t)\|_{H^{s-1}}$ and its spatial
derivatives can then be estimated using (4.8). Next apply the
argument of the last paragraph of section 3 to estimate the last term
in (5.3). Finally, the equation for $\dot\phi_m$ implies an estimate
for it in terms of $a_n$.                           \halmos

\vskip .5cm
I claim that the inequalities (5.5) and (5.6) imply the boundedness of
$a_n$ and $b_n$ uniformly in $n$ on an appropriate time interval. To
see this, first let $K_1$ be a constant which is greater than $C$ and
the supremum of $a_0(t)+b_0(t)$. Next choose $\lambda_0$ and $T$ so that
if $\lambda$ and $t$ are restricted to lie in the intervals
$(0,\lambda_0\rbrack$ and $\lbrack0,T)$ respectively the quantity
$\|\gamma_{ab}-\delta_{ab}\|_{K^{s-1}}+\|\gamma_{ab}-\delta_{ab}\|_p$
is small enough whenever $a_n$ is
less than $K_1$ so that the equation for $\phi_{n+1}$ can be solved and
(5.5) and (5.6) hold. Let $K_2$ be a bound for
$D^\prime$ under the condition that $\sup a_n\le K_1$. Let $K_3$ be
a bound for $D$ under the conditions that $\sup a_n\le K_1$ and
$\sup b_n\le K_2$. Reduce the size of $T$ if necessary so that
$$(C+K_2K_3T)e^{K_3T}\le K_1.\eqno(5.7)$$
By induction $a_n(t)\le K_1$ and $b_n(t)\le K_2$ for all
$t\in\lbrack0,T)$
and all $n$. Thus the iteration is bounded as claimed. This bounded
iteration is what is needed to show the existence of a regular solution
of the equations, as will now be shown.

\noindent
{\bf Definition} A solution $(g_{ab}(\lambda),k_{ab}(\lambda),
\phi(\lambda),f(\lambda))$ of the Vlasov-Einstein system is called
{\it regular of order $s$} if for each $\lambda$

\noindent
(i) $g_{ab}(\lambda)-\lambda\delta_{ab}$ and its time derivative
belong to $L^\infty(\lbrack 0,T),K^s(\R^3))$.

\noindent
(ii) $k_{ab}(\lambda)$ belongs to $L^\infty(\lbrack0,T), H^s(\R^3))$
and $\d_t k_{ab}(\lambda)$
belongs to $L^\infty(\lbrack0,T), H^{s-1}(\R^3))$

\noindent
(iii) $\phi(\lambda)$ and its time derivative belong to
$L^\infty(\lbrack0,T), K^{s+1}(\R^3))$

\noindent
(iv) $f(\lambda)$ and its time derivative belong to
$L^\infty(\lbrack0,T), H^{s-1}(\R^6))$
\vskip .2cm\noindent

The boundedness of the iteration shows that $\gamma_n$
is bounded in $C^1(\lbrack0,T), K^s(\R^3))$ for each fixed $\lambda$.
In particular $\gamma_n$ is bounded in
$L^\infty(\lbrack 0,T)\times\R^3)$
and $\nabla\gamma_n$ is bounded in $L^\infty(\lbrack 0,T),
H^{s-1}(\R^3))$. Each of these spaces is the dual of a Banach space and
so any bounded sequence has a subsequence which
converges in the weak$^*$ topology. By passing to a subsequence again it
can be ensured that the time derivatives of the $\gamma_n$ converge in
that sense. In the following we will always use the same notation for a
subsequence as for the original sequence. Now let $R>0$ be a real number
and let $\gamma_n(R)$ denote the restriction of $\gamma_n$ to
the closed ball $B_R$ of radius $R$ about the origin in $\R^3$. A
similar notation will be used for other objects defined on $\R^3$. The
sequence $\{\gamma_n(R)\}$ is bounded in $C^1(\lbrack 0,T),H^s(B_R))$.
Furthermore the embedding of $H^s(B_R)$ into $H^{s-1}(B_R)$ is compact.
Using the vector-valued Ascoli theorem\ref9\ it can be concluded that
there is a subsequence which converges strongly in $C^0(\lbrack0,T),
H^{s-1}(B_R))$. A diagonal argument shows that the a subsequence can
even be chosen so that this is true for all $R$. By passing repeatedly
to a subsequence and using the same argument it is possible to obtain
similar convergence statements for $k_n$, $\nabla\phi_n$. For $f_n$
it is even possible to get strong convergence in
$C^0(\lbrack0,T), H^{s-2}(\R^6))$ since the supports of all $f_n$ are
known to be contained in a fixed compact subset of $R^6$. In the
end we obtain a sequence which converges weakly in a space defined
globally on $R^3$ (or $R^6$ in the case of $f_n$) and strongly in
spaces defined on compact subsets. The weak convergence ensures
that the limit of the this sequence belongs to the spaces specified
in the definition of a regular solution while the strong convergence
ensures that it does define a solution of the Vlasov-Einstein system.

\vskip .5cm
\noindent
{\bf 6. Convergence to the Newtonian limit}

First it is necessary to get hold of the functions $f_N$ and $U$ which
occur in the statement of Theorem 1.1. Now $f(\lambda)$ is bounded in
$C^1(\lbrack0,T),H^{s-1}(\R^6))$. Choose a sequence $\{\lambda_n\}$
converging to zero. Let $f_n=f(\lambda_n)$. (This should not be
confused with the sequence of iterates used in the last section.) The
sequence $\{f_n\}$ is bounded in $C^1(\lbrack0,T),H^{s-1}(\R^6))$ and
hence, the Ascoli theorem for vector-valued functions, has a
subsequence which converges strongly in
$C^0(\lbrack0,T),H^{s-2}(\R^6))$. Call the limit $f_N$. Next consider
the sequence $\{\lambda_n^{-1}(1-\phi_n)\}$ where
$\phi_n=\phi(\lambda_n)$. This is bounded in
$C^1(\lbrack0,T)\times\R^3))$ and the sequence
$\{\lambda_n^{-1}\nabla\phi_n\}$ is bounded in
$C^1(\lbrack0,T),H^s(\R^3))$. By the type of argument used in the
previous
section it can be seen that, after passing to a subsequence,
$\{\lambda_n^{-1}(1-\phi_n)\}$ converges uniformly on compact subsets
to a limit, which will be denoted by $U$, and that for any $R>0$ the
restriction of $\lambda_n^{-1}\nabla\phi_n$ to $B_R$ converges in
$C^0(\lbrack0,T), H^{s-1}(B_R))$ to the restriction of $\nabla U$.
Moreover, using the weak${}^*$ convergence argument, $U$ is bounded
and $\nabla U$ is in $L^\infty(\lbrack0,T),H^s(\R^3))$. In the same way
it can be concluded that there is a sequence $\{\lambda_n\}$ converging
to zero such that $\gamma_n=\gamma(\lambda_n)$ and
$\kappa_n=\kappa(\lambda_n)$ converge in suitable function spaces as
$n\to\infty$. From equation (4.4) it can be seen that the Ricci tensor
of $\gamma_n$ tends to zero as $n\to\infty$. Hence the limiting metric
is flat. Consider now the time derivative of the connection. This is
given by equation (4.8). We know already that $\nabla\phi$ is
$O(\lambda)$ and that $\nabla\kappa$ is $O(\lambda^{1/2})$ in certain
spaces. Hence equation (4.8) implies that $\dot\Gamma^a_{bc}$ is
$O(\lambda^{1/2})$. Integrating in time shows that the Christoffel
symbols of $\gamma_{ab}$ are $O(\lambda^{1/2})$. Now the partial
derivatives of $\gamma_{ab}$ with respect to the spatial coordinates can
be written in terms of these Christoffel symbols and $\gamma_{ab}$
itself as in (4.14) and it follows that the partial derivatives of
$\gamma_{ab}$ must be $O(\lambda^{1/2})$. In particular the partial
derivatives of the limit of the $\gamma_n$ vanish and so this limiting
metric must in fact be given by $\delta_{ab}$. As a consequence the
limit of the $\kappa_n$ must also vanish. The characteristics of the
Vlasov equation converge uniformly to those of the non-relativistic
Vlasov equation with force term $\nabla U$ along the sequence
$\lambda_n$. Hence $f_N$ must coincide with the unique solution of
the latter equation with the $C^1$ initial datum $f_N^0$. Passing
to the limit in (2.30) and (3.1) then shows that $f_N$ and $U$
satisfy the Vlasov-Poisson system with initial datum $f_N^0$.
Unfortunately the convergence statements derived up to now are not
enough to ensure that the function $U$ satisfies the standard
boundary condition that $U\to 0$ as $r\to\infty$. On the other hand
it is known that $U$ is bounded so that it can be made to satisfy the
boundary condition by adding to it an appropriate constant. It will
be supposed from now on that this alteration has been made so that
$U$ vanishes at infinity. Since the solution of the Vlasov-Poisson
system with a given $C^1$ initial datum is unique, it follows that
for every sequence $\{\lambda_n\}$ the sequence $\{\lambda^{-1}\nabla
\phi_n, f_n\}$ has a subsequence tending to the same limit, namely
$(\nabla U,f_N)$. Hence the restriction of $\lambda^{-1}\nabla\phi$ to
any ball of radius $R$ converges to $\nabla U$ in $H^{s-1}$ and $f_n$
converges to $f$ in $H^{s-2}$. It is also possible to show weak${}^*$
convergence globally in $\R^3$.

The convergence statements obtained up to now can be improved by
estimating the difference between the Newtonian and relativistic
solutions.
$$\eqalign{
&\Delta_f(\lambda^{-1}(\phi-1)+U)=-(\gamma^{ab}-\delta^{ab})\d_a\d_b
(\lambda^{-1}\phi)-\gamma^{ab}\Gamma^c_{ab}(\lambda^{-1}\d_c\phi)
\cr
&+\lbrack |\kappa|^2+4\pi\phi^{-2}T_{00}-4\pi T_{00}(0)
+4\pi T\rbrack\phi}\eqno(6.1)$$
Hence
$$\eqalign{
&\|\lambda^{-1}(\phi-1)+U\|_{K^s}\le
C(\|\gamma_{ab}-\delta_{ab}\|_{K^{s-1}}
+\|\kappa_{ab}\|_{H^{s-2}}       \cr
&+\|T_{00}-T_{00}(0)\|_{H^{s-2}}+\lambda)}\eqno(6.2)$$
Differentiating (6.1) gives the estimate
$$\eqalign{
&\|\lambda^{-1}\dot\phi+\dot U\|_{K^s}\le
C(\|\gamma_{ab}-\delta_{ab}\|_{K^{s-1}}+\|\d_t\gamma_{ab}\|_{H^{s-1}}
\cr
&+\|\kappa_{ab}\|_{H^{s-2}}+\|\d_t\kappa_{ab}\|_{H^{s-2}}
+\|\dot T_{00}-\dot T_{00}\|_{H^{s-2}}
+\lambda)}\eqno(6.3)$$
Using the definition of $T_{00}$ it can be shown straightforwardly that
the terms involving $T_{00}$ in (6.2) and (6.3) can be replaced in these
inequalities by expressions involving $f-f_N$.
$$\eqalign{
&\|\lambda^{-1}(\phi-1)+U\|_{K^s}\le
C(\|\gamma_{ab}-\delta_{ab}\|_{K^{s-1}}+\|\kappa_{ab}\|_{H^{s-2}} \cr
&\qquad+\|f-f_N\|_{H^{s-2}}+\lambda)}\eqno(6.4)$$
The inequality (6.3) can be modified similarly. Subtracting the Vlasov
equation for $f_N$ from that for $f$ gives the estimate
$$\eqalign{
&\|f(t)-f_N(t)\|_{H^{s-2}}^2\le\|f(0)-f_N(0)\|_{H^{s-2}}^2
+C\int_0^t(\|f(t^\prime)-f_N(t^\prime)\|_{H^{s-2}}^2       \cr
&+\|\gamma_{ab}(t^\prime)-\delta_{ab}\|_{K^{s-2}}^2
+\|\lambda^{-1}\nabla_c\phi(t^\prime)+
\nabla_c U(t^\prime)\|_{H^{s-1}}^2+\lambda^2)dt^\prime}\eqno(6.5)$$
and an analogous estimate for $\|\dot f(t)-\dot f_N(t)\|_{H^{s-2}}$.
Equation (2.29) implies an estimate of the form
$$\|\gamma_{ab}(t)-\gamma_{ab}(0)\|_{H^{s-1}}^2\le C\int_0^t
\|\gamma_{ab}(t^\prime)-\gamma_{ab}(0)\|_{H^{s-1}}^2
+\|\kappa_{ab}(t^\prime)\|_{H^{s-1}}^2 dt^\prime\eqno(6.6)$$
To close the argument and obtain a useful differential inequality,
it remains to estimate $\kappa_{ab}$. This can be done by examining
carefully the third term in (4.12). This is mostly routine but there
are two expressions which require particular care and these will be
handled explicitly here. The first, which arises from $N_{ab}$ is:
$$\eqalign{
&\left|\int_{\R^3}\lambda^{-1}\gamma^{ac}\gamma^{bd}\phi^{-2}\nabla_c
\nabla_d\dot\phi\d_t\kappa_{ab} dV_\gamma\right|          \cr
&\le\left|\int_{\R^3}\lambda^{-1}\gamma^{ac}\gamma^{bd}
\phi^{-2}\nabla_d
\dot\phi\nabla_c(\d_t\kappa_{ab}) dV_\gamma\right|+\ldots  \cr
&\le C\lambda^{1/2}(\|\lambda^{-1}\nabla_d\dot\phi\|_2^2+
\|\lambda^{-1/2}\nabla_c(\d_t\kappa_{ab})\|^2_2)+\ldots}$$
The other, which arises from $\tau_{ab}$, is
$|\int_{\R^3}\gamma^{ac}\gamma^{bd}\phi^{-3}\dot T_{00}\gamma_{cd}
\d_t\kappa_{ab} dV_\gamma |$
To estimate this, first note that $\dot T^{00}-\dot T^{00}(0)$ can
be controlled straightforwardly so that $\dot T^{00}$ can be replaced
without loss of generality by $\dot T^{00}(0)$. Now the fact can be
used that $\dot T^{00}(0)=-\d_a T^{0a}(0)$. After this substitution
has been made it suffices to integrate by parts in space.

The quantity $\d_t\gamma_{ab}$ satisfies $\|\d_t\gamma_{ab}\|_{H^{s-1}}
\le \|\kappa_{ab}\|_{H^{s-1}}$. Furthermore:
$$\|\gamma_{ab}(t)-\delta_{ab}\|_{K^{s-1}}\le
\|\gamma_{ab}(t)-\gamma_{ab}(0)\|_{H^{s-1}}
+\|\gamma_{ab}(0)-\delta_{ab}\|_{H^{s-1}_\delta}\eqno(6.7)$$
Define
$$\eqalign{
a(t)&=\sup_\lambda\{\|\gamma_{ab}(t)-\gamma_{ab}(0)\|_{H^{s-1}}^2+
\|\kappa_{ab}(t)\|_{H^{s-1}}^2+\|\d_t\kappa_{ab}(t)\|_{H^{s-2}}^2
\cr
&+\|\lambda^{-1/2}\nabla\kappa_{ab}(t)\|_{H^{s-2}}^2
+\|f(t)-f_N(t)\|_{H^{s-2}}^2
+\|\d_t f(t)-\d_t f_N(t)\|_{H^{s-2}}^2\}}\eqno(6.8)$$
Then the above estimates show that
$$a(t)\le C(\lambda^{1/2}+\int_0^t a(t^\prime)dt^\prime)\eqno(6.9)$$
It follows using Gronwall's inequality that $a(t)=O(\lambda^{1/2})$.

The meaning of the order symbols in the statement of Theorem 1 can now
be explained. They refer to the function spaces obtained from those
occurring in the definition of a regular solution of the
Vlasov-Einstein system by replacing $s$ by $s-1$. In other words,
convergence is obtained in a space involving one less derivative
than that where the existence of the solution has been obtained.
Given this definition, the conclusions (i)-(vi) of Theorem 1.1 follow
from (6.9), (6.7) and (6.4).

\vskip .5cm
\noindent
{\bf 7. Solution of the constraints}

In terms of $\gamma_{ab}$ and $\kappa_{ab}$ these take the form
$$\eqalignno{
R_\gamma-\lambda|\kappa|^2&=16\pi\lambda\phi^{-2}T_{00},&(7.1)    \cr
\gamma^{ab}\nabla_a\kappa_{bc}&=-8\pi\phi^{-1}T_{0c}&(7.2)}$$
for a maximal hypersurface. Let $\mu=\phi^{-2}T_{00}$ and
$J_a=\phi^{-1}T_{0a}$. Note that when $\mu$ and $J_a$ are expressed
in terms of $f$ and $\gamma_{ab}$ using (3.7) then it is seen that
they do not explicitly depend on $\phi$. To solve these equations we
start with the following $\lambda$-dependent objects:

\noindent
(i) a function $\tilde\mu$ of the form $\tilde\mu(0)
+O(\lambda^{1/2})$

\noindent
(ii) a vector $\tilde J_a$ which is $O(\lambda)$ as $\lambda\to 0$

\noindent
(iii) a metric $\tilde\gamma_{ab}$ satisfying
$\tilde\gamma_{ab}=\delta_{ab}+O(\lambda^{3/2})$

\noindent
(iv) a traceless symmetric tensor $\tilde\kappa_{ab}$ satisfying
$\tilde\kappa_{ab}=O(\lambda^{1/2})$ and $\tilde\gamma^{ab}
\tilde\nabla_a\tilde\kappa_{bc}=-8\pi\tilde J_c$

\noindent
It is assumed that $\tilde\mu\in H^s(\R^3)$,
$\tilde J_a\in H^s(\R^3)$
and $\tilde\gamma_{ab}-\delta_{ab}\in H^s_\delta(\R^3)$ for some
$\delta$ in the interval $(-1,-1/2)$ and that all these objects
depend continuously on $\lambda$ in the given spaces. The $O$-symbols
here also refer to those spaces. Furthermore, it
is assumed that $\tilde\mu$ and $\tilde J_a$ are compactly
supported. Tensors $\tilde\kappa_{ab}$ of the type required in (iv)
can be contructed in a standard way using the York decomposition.
{}From these objects it is possible to determine initial data
for the Einstein equations by solving the equation\ref3
$$\Delta_{\tilde\gamma}\psi+(1/8)(-R_{\tilde\gamma}\psi+\lambda
|\tilde\kappa|^2\psi^{-7})+2\pi\lambda\tilde\mu\psi^{-3}=0
\eqno(7.3)$$
and definining $\gamma=\psi^4\tilde\gamma$, $\kappa=\psi^{-2}\tilde
\kappa$, $\phi^{-1}T_{0a}=J_a=\psi^{-6}\tilde J_a$ and $\phi^{-2}T_{00}
=\mu=\psi^{-8}\tilde\mu$. It is well known that equation (7.3) can be
solved for a unique $\psi$ which tends to 1 at infinity and which has
the property that $\psi-1$ belongs to a weighted Sobolev space provided
that $\tilde\gamma_{ab}$ is close to $\delta_{ab}$ in a weighted
Sobolev space. Because of assumption (iii) above this will be satisfied
when $\lambda$ is close to zero. Note that the condition $\delta>-1$
implies that if $s\ge2$ the space $H^s_{\delta}(\R^3)$ is continuously
embedded in $L^p(\R^3)$ for some $p<6$. Let $\psi_1$ be the solution of
the equation $\Delta_f\psi_1=-2\pi\tilde\mu(0)$ which tends to zero at
infinity. Then since $U$ satisfies the Poisson equation and $\psi$ is
identically one when $\lambda=0$ it follows that $\psi_1=(1/2)U$.
Now a comparison with (7.3) shows that
$$\Delta_f(\psi-\lambda\psi_1)=O(\lambda^{3/2})\eqno(7.4)$$
Hence $\psi=1+\lambda\psi_1+O(\lambda^{3/2})$. This can be used to
calculate the Ricci tensor of $\gamma_{ij}$ to order $\lambda$ using the
formula for conformal transformations. The result is that the
contributions of order $\lambda$ on the right hand side of (2.4) cancel
so that assumption (iii) of Theorem 1.1 is satisfied by the given
initial data. Now the significance of this assumption can be discussed.
In \ref{21} it was shown that a necessary condition for the existence of
a regular Newtonian limit is that the coefficient of $\lambda^2$ in the
expansion of the spatial part of the metric, considered as a linearised
metric, has vanishing linearised Bach tensor. This means that it
satisfies the linearised version of the condition of conformal flatness.
Another way of expressing this would be to say that the spatial metric
is conformally flat to second order in $\lambda$. It was in order to
ensure that this condition was satisfied that $\tilde\gamma$ was
chosen to be flat up to first order in $\lambda$. The computation
which has just been done shows that the role of this condition in
the present paper is to ensure the correct behaviour of
$\d_t\kappa_{ab}$ for $\lambda\to 0$ on the initial hypersurface
in order to make the iteration work.

To get a complete initial data set for the reduced equations it is
still necessary to describe how the initial values for $f$ are to be
obtained. This is done as follows. Suppose that a test metric
$\tilde\gamma_{ab}$ has been chosen as in (iii) above. Let $\tilde f$
be a non-negative real-valued function of compact support on the mass
shell defined by $\tilde\gamma_{ab}$. Define $\tilde\mu$ and
$\tilde J_a$ in terms of $\tilde f$ and $\tilde\gamma_{ab}$ in
the same way as $\mu=\phi^{-2}T_{00}$ and $J_a=\phi^{-1}T_{0a}$ are
defined in terms of $f$ and $\gamma_{ab}$ (i.e by (3.7)). Now construct
the quantities $\gamma_{ab}$, $\kappa_{ab}$, $T_{0a}$ and $T_{00}$ as
described earlier in this section. Let $\tilde p^a=\psi^2 p^a$ and
define
$f(p^a)=\psi^{-8}\tilde f(\tilde p^a)$. It then follows that $T_{00}$
and $T_{0a}$ arise from $f$ via the equation (3.7) and a full initial
data set for the Vlasov-Einstein system is obtained.

\vskip .5cm
\noindent
{\bf Appendix 1. Some elliptic theory}

The discussion here will be limited to results which are not standard.
For background material on the Poisson integral see \ref7. In this
appendix $\Delta$ always denotes the Laplacian of the standard flat
metric on $\R^3$. For any function $\phi$ on $\R^3$ and a positive real
number $\epsilon$ let $\phi_\epsilon(x)=\phi(\epsilon x)$. If $f$ is a
continuous function belonging to $L^p(\R^3)\cap L^\infty(\R^3)$ for some
$p<3/2$ then it possesses a Newtonian potential $u$ which is $C^1$. This
is a solution of the Poisson equation $\Delta u=f$ in the sense of
distributions and can be obtained as the limit for $\epsilon\to 0$ of
the Newtonian potentials of the compactly supported functions
$\phi_\epsilon f$ where $\phi$ is any $C^\infty$ function of compact
support which takes the value 1 in a neighbourhood of the origin. The
Newtonian potentials of the functions $\phi_\epsilon f$ can be
represented by the familiar Poisson integral. They tend to zero at
infinity. Pick a sequence $\epsilon_n$ tending to zero and let
$f_n=\phi_{\epsilon_n} f$. Then the $f_n$ are bounded in
$L^\infty(\R^3)$ and converge to $f$ in $L^p(\R^3)$. Let $u_n$ denote
the Newtonian potential of $f_n$. A useful estimate for the Poisson
integral will now be derived (cf. \ref2). For any $R>0$:
$$\eqalign{\int_{\R^3}{f(y)\over |x-y|} dy&=\int_{|x-y|<R}{f(y)\over
|x-y|} dy+
\int_{|x-y|>R}{f(y)\over |x-y|} dy             \cr
&\le C(\|f\|_\infty R^2+\|f\|_p R^{(3-q)/q}).}$$
Here it has been assumed that the conjugate exponent $q$ of $p$ is
greater
than 3, which implies that $p<3/2$. Putting
$R=(\|f\|_p/\|f\|_\infty)^{p/3}$
gives the estimate
$$\|u_n\|_\infty\le C\|f_n\|_p^{2p/3}\|f_n\|_\infty^{1-2p/3},\ \ \
p<3/2.
\eqno({\rm A}1)$$
The derivatives of $u_n$ can be estimated similarly.
$$\|\nabla u_n\|_\infty\le C\|f_n\|_p^{p/3}\|f_n\|_\infty^{1-p/3},\ \ \
p<3.
\eqno({\rm A}2)$$
These estimates show that $u_n$ and its first derivatives converge
uniformly to $u$ and its first derivatives. In particular $u$ tends to
zero as $|x|\to\infty$. Now all the functions $u_n$ have the property
that $u_n(x)=O(|x|^{-1})$ and $\nabla u_n(x)=O(|x|^{-2})$ as
$|x|\to\infty$. Hence a partial integration shows that
$$\|\nabla u_n\|_2^2\le\|u_n\|_\infty\|f_n\|_1.
\eqno({\rm A}3)$$
If $f\in L^1(R^3)$ it follows that $\nabla u_n$ is a bounded sequence in
$L^2(R^3)$. Since this sequence also converges pointwise it follows from
Fatou's lemma that $\nabla u$ is in $L^2(R^3)$ and that it satisfies the
equivalent of (A3). Putting this together with (A1) gives
$$\|\nabla u\|_2\le C\|f\|_\infty^{1/6}\|f\|_1^{5/6}.
\eqno({\rm A}4)$$
Another estimate which is satisfied by $u$ when $f\in H^s(\R^3)$ is
$$\|\d_a\d_b u\|_{H^s}\le \|f\|_{H^s}.\eqno({\rm A}5)$$
It has now been shown that if $f\in L^1(\R^3)\cap H^s(\R^3)$
then its Newtonian potential $u$ belongs to $L^\infty(\R^3)$ and
$\d_a u$ belongs to $H^{s+1}(\R^3)$. In the following another
related result will be required. Let $H^{ab}$ be a tensor on $\R^3$
whose components belong to $L^p(\R^3)\cap L^\infty(\R^3)$ for some
$p<6$ and which has the property that $\d_c H^{ab}$ belongs to
$H^{s-1}(\R^3)$. Let $h$ be a function which is bounded and whose
first derivatives are in $H^s(\R^3)$. Consider the equation
$\Delta u=H^{ab}\d_a\d_bh$. By H\"older's inequality the expression
on the right hand side of this equation belongs to $L^q(\R^3)$ for
some $q<3/2$. Thus it can be concluded from the above discussion
that $u$ is in $L^\infty(\R^3)$ and an estimate for its norm in
that space follows from (A1). Now let $H^{ab}_n=\phi_{\epsilon_n}
H^{ab}$ and let $u_n$ be the solution of $\Delta u_n=H^{ab}_n\d_a\d_bh$.
$$\eqalign{\|\nabla u_n\|_2^2&=-\int_{\R^3} u_nH_n^{ab}\d_a\d_b h  \cr
&=-\int_{\R^3} (\d_a u_nH_n^{ab}\d_b h+u_n\d_aH_n^{ab}\d_b h).}$$
Thus the $L^2$ norm of $\nabla u_n$ is bounded and it is possible to
argue as above that $\nabla u$ is in $L^2(\R^3)$ and that its norm
can be estimated in terms of $\|H^{ab}\|_\infty$, $\|\d_a H^{ab}\|_2$
and $\|\d_b h\|_2$.

\noindent
{\bf Lemma A1} {\it Let $\gamma_{ab}$ be a Riemannian metric such
that $\gamma_{ab}-\delta_{ab}\in L^p(\R^3)\cap K^s(\R^3)$ for
some $p<6$ and some $s\ge 3$ and
let $\Gamma^a$ be its contracted Christoffel symbols. Let $\rho$
be a function belonging to $L^1(\R^3)\cap H^{s-1}(\R^3)$ and
$\lambda$ a non-negative real number. Then there exist positive
constants $C_1$ and $C_2$ such that if
$\|\gamma_{ab}-\delta_{ab}\|_p
+\|\gamma_{ab}-\delta_{ab}\|_{K^{s-1}}\le C_1$ and
$(\|\rho\|_1+\|\rho\|_{H^{s-1}})\lambda\le C_2$
then the equation
$$\Delta\phi=(\delta^{ab}-\gamma^{ab})\d_a\d_b\phi+\Gamma^a\d_a\phi
+\lambda\rho\phi\eqno({\rm A}6)$$
has a unique solution with the property that $\phi\to 1$ as $|x|\to
\infty$. Moreover for any $k\le s$ the solution satisfies an estimate of
the form
$$\|\lambda^{-1}(\phi-1)\|_\infty+\|\lambda^{-1}\nabla\phi\|_{H^k}\le C
\eqno({\rm A}7)$$
for a constant $C$ only depending on $C_1$, $C_2$ and $\|\gamma_{ab}-
\delta_{ab}\|_{K^s}$. }

\noindent
{\it Proof} Define an iteration by solving
$$\Delta\phi_{n+1}=(\delta^{ab}-\gamma^{ab})\d_a\d_b\phi_n+\Gamma^a\d_a
\phi_n+\lambda\rho\phi_n\eqno({\rm A}8)$$
with $\phi_{n+1}\to 1$ as $|x|\to\infty$ and $\phi_0=1$. Let
$q_n=\Gamma^a\d_a\phi+\lambda\rho\phi$. If $\phi_n\in L^\infty(\R^3)$
and $\nabla\phi_n\in H^s(\R^3)$ then $q_n\in L^1(\R^3)\cap
H^{s-1}(\R^3)$
and so by the above remarks concerning the Poisson equation the solution
$\phi_{n+1}$ exists. Furthermore
$$\|q_n\|_1+\|q_n\|_{H^{k-1}}\le C(\|\gamma_{ab}-\delta_{ab}\|_{K^k}+
\lambda(\|\rho\|_1+\|\rho\|_{H^{k-1}}))(\|\phi_n\|_\infty
+\|\nabla\phi_n\|_{H^k})\eqno({\rm A}9)$$
for any $k$ with $3\le k\le s$. Thus there exist constants $C_1$ and
$C_2$ such that if the inequalities
$\|\gamma_{ab}-\delta_{ab}\|_{L^p}+\|\gamma_{ab}-\delta_{ab}\|_{K^3}
\le C_1$ and
$(\|\rho\|_1+\|\rho\|_{H^{s-1}})\lambda\le C_2$ hold then
$$\|\phi_{n+1}-\phi_n\|_\infty+\|\nabla\phi_{n+1}-\nabla\phi_n\|_{H^3}
\le K(\|\phi_n-\phi_{n-1}\|_\infty
+\|\nabla\phi_n-\nabla\phi_{n-1}\|_{H^3})
\eqno({\rm A}10)$$
for some $K<1$. It follows that $\{\phi_n\}$ and $\{\nabla\phi_n\}$ are
Cauchy sequences in $L^\infty(\R^3)$ and $H^3(\R^3)$ respectively and
hence $\phi_n$ converges to a solution of (A6) with $\phi\to 1$
as $|x|\to \infty$. This solution satisfies
$$\|\phi\|_\infty+\|\nabla\phi\|_{H^3}\le C\eqno({\rm A} 11)$$
for some $C$ depending only on $C_1$ and $C_2$. Dividing (A6)
by $\lambda$ gives
$$\Delta(\lambda^{-1}\phi)=(\delta^{ab}-\gamma^{ab})\d_a\d_b
(\lambda^{-1}\phi)+\Gamma^a\d_a(\lambda^{-1}\phi)
+\rho\phi\eqno({\rm A}12)$$
Estimating the quantity $\lambda^{-1}\phi_n$ in the same way as $\phi_n$
was estimated above shows that $\{\lambda^{-1}(\phi_n-1)\}$ and
$\{\lambda^{-1}\nabla\phi_n\}$ are Cauchy sequences in $L^\infty(\R^3)$
and $H^3(\R^3)$ respectively and that (A7) holds for $k=3$.
To see the that it is true for any $k\le s$, use the information we
already
have in the right hand side of (A12). Using the estimates for the
Poisson equation stated at the beginning of this appendix then shows
that if the hypotheses of the lemma hold and if $s\ge4$ we obtain (A7)
for $k=4$ after possibly reducing the size of $C_1$. This process can be
repeated until $k=s$.    \halmos

\vskip .5cm
\noindent
{\bf Appendix 2. Estimates for modified Sobolev spaces}

This appendix is concerned with proving some useful estimates for the
modified Sobolev spaces $K^s(\R^3)$ introduced in section 3. Recall that
$$\|f\|_{K^s}=\|f\|_\infty+\|\nabla f\|_{H^{s-1}}.\eqno({\rm A}13)$$
The results to be proved are analogues of the results (3.2) and (3.3)
for functions belonging to ordinary Sobolev spaces and are proved in
a similar way. Estimates of this type are discussed in \ref{14},
\ref{16}\ and \ref{23}. First analogues of (3.2) will be discussed.
Suppose that $f,g$ belong to $K^s(\R^3)$ for some $s\ge 2$. Note first
the obvious fact that $\|fg\|_\infty\le\|f\|_\infty\|g\|_\infty$.
If $\alpha$ is a multi-index with $1\le |\alpha|\le s$ then
$$D^\alpha(fg)=D^\alpha f g+fD^\alpha g+\sum{\alpha!\over\beta!\gamma!}
D^\beta fD^\gamma g.\eqno({\rm A}14)$$
The sum is taken over all multi-indices $\beta$ and $\gamma$ with
$|\beta|+|\gamma|=|\alpha|$ and $\max (|\beta|, |\gamma|)<|\alpha|$.
The first estimate we wish to prove is that
$$\|D^\alpha(fg)\|_2\le C\|f\|_{K^s}\|g\|_{K^s}\eqno({\rm A}15)$$
for some constant $C$. It is elementary to estimate the first two terms
in (A14) by the right hand side of (A15) and so we can concentrate on
the third term. Consider one of the summands there. Suppose first that
either $\beta$ or $\gamma$ is less than $s-3/2$. Without loss of
generality we can assume that it is $\beta$. Then
$$\|D^\beta fD^\gamma g\|_2\le\|D^\beta f\|_\infty\|D^\gamma g\|_2
\le C\|D^\beta f\|_{H^2}\|D^\gamma g\|_2\eqno({\rm A}16)$$
If on the other hand both $\beta$ and $\gamma$ are greater than $s-3/2$
then it can be concluded that $s<3$. The only estimate which
remains to be done to establish (A15) is
$$\|DfDg\|_2\le\|Df\|_4\|Dg\|_4\le\|Df\|_{H^1}\|Dg\|_{H^1},\eqno({\rm A}
17)$$
where the H\"older and Sobolev inequalities have been used. The
inequality (A15) shows that multiplication is a continuous mapping
from $K^s(\R^3)$ to itself. This is a statement of a weaker type than
(3.2). A stronger result could presumably be obtained using the
Gagliardo-Nirenberg inequality but that will not be attempted here
since (A15) is sufficient for the applications in this paper. In a
similar way it can be shown that if $f\in H^s(\R^3)$ and
$g\in K^s(\R^3)$ for $s\ge2$ then $fg\in H^s(\R^3)$ and
$$\|fg\|_{H^s}\le\|f\|_{H^s}\|g\|_{K^s}.\eqno({\rm A} 18)$$
An estimate related to (3.3) can also be obtained if $s\ge 3$, namely
$$\|D^\alpha(fg)-fD^\alpha g\|_2\le C\|f\|_{K^s}\|g\|_{K^{s-1}},\ \ \
1\le|\alpha|\le s\eqno({\rm A} 19)$$

The final estimate which is required concerns composition with a $C^r$
function. Suppose then that $U$ is an open interval in $\R$ and
$F:U\to\R$ a $C^r$ function whose derivatives up to order $r$ are
bounded on $U$. Let $f$ be a function in $K^s(\R^3)$, where
$2\le s\le r$, whose range is contained in $U$. Now
$$D^\alpha(F(f))=\sum C_{r\alpha_1\ldots\alpha_l}{d^k F\over df^k}
D^{\alpha_1}f\ldots D^{\alpha_l}f\eqno({\rm A}20)$$
where $1\le k\le |\alpha |$, $|\alpha_1|+\ldots+|\alpha_l|=|\alpha|$
and $|\alpha_i|\ge1$ for all $i$. By an argument similar to those
given above it is seen that the terms on the right hand side of (A20)
can be estimated straightforwardly unless $s=2$. In that exceptional
case the embedding $H^1(\R^3)\to L^4(R^3)$ can be used. There results
the estimate
$$\|F(f)\|_{K^s}\le C\|F\|_{C^r}\|f\|_{K^s}^s.\eqno({\rm A}21)$$
In this paper (A21) is only needed to estimate $\phi^{-1}$ and the
inverse metric $\gamma^{ab}$. In both cases it is applied to the
function $F(f)=1/f$ with $U=(c,\infty)$ for some $c>0$.

\vskip 10pt
\noindent
{\it Acknowledgements} I thank D. Christodoulou for suggesting the
argument involving integration by parts in time used in the proof of
Lemma 2.1 and again in Sect. 4. I am also grateful to G. Rein for
helpful comments.

\vskip .5cm
\noindent
{\bf References}

\noindent
1.Asano, K., Ukai, S.: On the Vlasov-Poisson limit of the Vlasov-Maxwell
equation. In Nishida,T., Mimura, M., Fujii, H. (eds.) Patterns and
Waves. North Holland, Amsterdam 1986
\next
2.Batt, J.: Global symmetric solutions of the initial value problem of
stellar dynamics. J. Diff. Eq. {\bf 25}, 342-364 (1977)
\next
3.Cantor, M.: A necessary and sufficient condition for York data to
specify an asymptotically flat spacetime. J. Math. Phys. {\bf 20},
1741-1744 (1979)
\next
4.Cartan, E.: Sur les vari\'et\'es \`a connexion affine et la th\'eorie
de la relativit\'e g\'en\'eralis\'ee. Ann. Sci. Ecole Norm. Sup.
{\bf 39}, 325-412 (1922); {\bf 41}, 1-25 (1924)
\next
5.Choquet-Bruhat, Y.: Probl\`eme de Cauchy pour le syst\`eme integro
differentiel d'Einstein-Liouville. Ann. Inst. Fourier (Grenoble)
{\bf 21}, 181-201 (1971)
\next
6.Christodoulou, D., Klainerman, S.: The global nonlinear stability of
the Minkowski space. Preprint.
\next
7.Dautray, R., Lions, J.-L. Mathematical analysis and numerical
methods for science and technology. Vol 1. Springer, Berlin 1990
\next
8.Degond, P.: Local existence of solutions of the Vlasov-Maxwell
equations and convergence to the Vlasov-Poisson equation for infinite
light velocity. Math. Meth. Appl. Sci. {\bf 8}, 533-558 (1986)
\next
9.Dieudonn\'e, J.: Foundations of modern analysis. Academic Press,
New York 1969
\next
10.Ebin, D., Marsden, J.: Groups of diffeomorphisms and the motion
of an incompressible fluid. Ann. of Math. {\bf 92}, 102-163 (1970)
\next
11.Ehlers, J.: The Newtonian limit of general relativity. In Ferrarese,
G. (ed.) Classical mechanics and relativity: relationship and
consistency. Bibliopolis, Naples 1991
\next
12.Friedrichs, K.O.: Eine invariante Formulierung des Newtonschen
Gravitationsgesetzes und des Grenz\"uberganges vom Einsteinschen
zum Newtonschen Gesetz. Math. Ann. {\bf 98}, 566-575 (1927)
\next
13.Fritelli, S., Reula, O. On the Newtonian limit of general
relativity. Preprint MPA 630, Garching
\next
14.Klainerman, S.: Global existence for nonlinear wave equations.
Commun. Pure Appl. Math. {\bf 33}, 43-101 (1980)
\next
15.Lottermoser, M.: A convergent post-Newtonian approximation for the
constraint equations in general relativity. Ann. Inst. H. Poincar\'e
(Physique Th\'eorique) {\bf 57}, 279-317 (1992)
\next
16.Majda, A.: Compressible fluid flow and systems of conservation laws
in several space variables. Springer, New York 1984
\next
17.Marsden, J., Ebin, D.G., Fischer, A.E.: Diffeomorphism groups,
hydrodynamics and relativity. In Vanstone, J.R. Proc. 13th Biennial
Seminar of the Canadian Mathematical Congress. Canadian Mathematical
Society, Montreal 1972
\next
18.Rein G., Rendall, A.D.: Global existence of solutions of the
spherically symmetric Vlasov-Einstein system with small initial data.
Commun. Math. Phys. {\bf 150}, 561-583 (1992)
\next
19.Rein G., Rendall, A.D.: The Newtonian limit of the spherically
symmetric Vlasov-Einstein system. Commun. Math. Phys. {\bf 150},
585-591 (1992)
\next
20.Rendall, A.D.: The initial value problem for a class of general
relativistic fluid bodies. J. Math. Phys. {\bf 33}, 1047-1053 (1992)
\next
21.Rendall, A.D.: On the definition of post-Newtonian approximations.
Proc. R. Soc. Lond. {\bf 438}, 341-360 (1992)
\next
22.Schaeffer, J.: The classical limit of the relativistic Vlasov-Maxwell
system. Commun. Math. Phys. {\bf 104}, 403-421 (1986)
\next
23.Zeidler, E.: Nonlinear functional analysis and its applications.
Vol. 2. Springer, New York 1990
%}
\end